# Nonlinear Dynamics in Complexity Quantification

Amin Gasmi

1$^{er}$ Août 2022

Chaotic systems which are due to nonlinearity have attracted a great concern in the current world and chaotic models. Systems for a wide range of operation conditions have their application in almost all branches of engineering and science. In the history of chaotic studies and nonlinearity, many different but co-existent phases can be distinguished [1]. In the initial phase, chaos was considered as a deterministic regime which, most probably, was responsible for the variations that was regarded as noise and thus was being modeled as a stochastic process. In the second phase, it was of great significance to establish criteria for detecting chaotic dynamics and thus establishing dynamical invariants which were necessary in quantifying chaos. The third step which was to develop machine learning models which could learn the dynamics and chaos from the data of the strange attractor [2]. With respect to this aspect, various model structures were developed and investigated on their ability to detect chaos from a given set of data [3]. These model structures included radial basis functions, and local linear mapping among others. The third phase is currently being investigated together with other issues surrounding nonlinear dynamics and chaos, for instance in noise reduction and control, among other issues.

# Modeling of Nonlinear Dynamics

Nonlinear dynamic modeling is an essential aspect in the understanding of nonlinear systems. Modeling gives a illustration of a nonlinear system. In this section, we will consider embedding techniques of nonlinear modeling.

## Embedding Techniques of Modeling Nonlinear system

We consider an $n^{th}$ order dynamic system represented by the equation y = f (y). A set of n first order ordinary differential equations can best represent the system, with a state variable governing each equation. It therefore imply that a global dynamic system would have n times variables $\{y_1, y_2, ..., y_n\}$ with an n time series solution [4]. The above-mentioned n time series is obtained from the initial nth order system through a decomposition process. Similarly, given the n times series, the initial solution can be obtained by taking individual state variable as a reconstruction space's coordinate. Therefore it is possible to use n time series in the composition and reconstruction of the system solution and trajectory.

In actual practice, there are various problems encountered when utilizing this approach. The main challenge arises from the order of the system. The order is rarely known and even if an accurate approximation of the variable exists, the number of measurements is not as large as the value n. For instance, if we consider the atmosphere as a system of a high order (i.e., n is too large), meteorological (weather forecasting) stations and monitoring departments usually measure only a very limited number of variables in making climatic and weather conditions' predictions. Mathematically, we can describe this approach by considering the action of measuring function $h(y): \square^n \to \square$ which is assumed to be operating in the entire phase space or state, but the function yields a scalar which is referred to as measured variable.

Naturally, a question arises from this state on the possibility of reconstructing a trajectory and a solution of f from such a scalar measurement h (y), given $f: \square^n \to \square^n$ and $h(y): \square^n \to \square$. This question, however, has an affirmative answer only when certain conditions are met. In state space (phase space), the steady space dynamic of a dynamic system can be represented by geometrical figures termed attractors. While a stable linear system has only one attractor kind, systems with nonlinear dynamic can be composed of many kinds of attractor which are more complicated such as tori, limit cycles, and strange attractors.

Hence, the presence of time series in the construction of phase (state) space of a dynamic system through embedding techniques makes it possible of using results from differential geometry and topology in the subsequent attractors analysis. The attractors are geometrical objects in the reconstructed space. Additionally, if the embedding is achieved successfully, reconstructed attractor topologically equals the original attractor and therefore considered to be diffeomorphic. The advantage of this techniques is that no matter the complexity of a dynamic system, there is the possibility of reconstructing the original phase space through embedding techniques [3]. Moreover, there is a possibility of estimating qualitative and quantitative invariants of the attractor, such as Lyapunov exponents, Poincare maps, and fractal dimension, directly from the reconstructed attractor. Usually, the reconstructed attractor must typologically equal to the original attractor.

Delay coordinates provide a convenient but exceptional method of phase spaces reconstruction from scalar measurements. However, there are other coordinates which can be used, such as derivative coordinates and singular value coordinates. Delay vector takes the form

$$y(k) = [y(k)\ y(k-\tau)\ ...\ y(k-(d_e-1))\tau]^T$$

Where

$$d_e = embedding\ dimension,\ \&\ \tau = delay\ time.$$

Therefore is is possible to the represent the function y (k) as a point in the $d_e$ – dimensional embedding space. A research conducted by [5], showed that embedding with $d_e$ greater than 2n gives accurate aswers generically resulting to a smooth map $f_T : \square^{d_e} \to \square$ such that;

$$y(k+T) = f_T(y(k))$$

For all cases where all k, T and τ have integer values (T= forecasting time), it is clear that the attractor which is reconstructed in $\square^{d_e}$ is diffeomorphic to the initial attractor, and thus the former retains the topological and dynamical characteristics of final attractor. When we consider a delay reconstruction, we must take a keen consideration on the choice of reconstruction parameters such as the the delay time and the embedding dimension which can be abbreviated as $\tau$ and $d_e$ respectively. This is because the quality of the embedded space is strongly affected by these two parameters. According to [5], many scientists have suggested that in various applications of nonlinear dynamics, it is more convenient and important to estimate $d_e$ and $\tau$ simultaneously, this is similar to estimating the embedding window which is defined as $(d_e - 1)\tau$. When the methods are compared, there is evidence that the optimum value of $\tau$ is a system identification problem is shorter than that of phase space reconstruction.

The second equation above can only hold if the condition is met. The condition is if dynamic invariants of the original system can be infered from a a single variable time series of. In estimating $f_T$, there are two primary methods which can be used, namely the global and local approximation techniques [3]. The local approach considers partitioning the embedding space into the neighborhood; $\{ui\}_{i=1}^{N_n}$ where appropriately the dynamics of the system can be described by a linear mapping:

$$g_T : \mathbb{R}^{d_e} \to \mathbb{R}$$

Such that

$$y(k+T) \approx g_{Ti}(y(k)) \text{ for } y(k) \in ui, \ i=1,\ldots,N_n$$

Various choices of $g_T$ have been proposed in various research works such as the use of linear polynomials which can be subjected to interpolation in obtaining an approximation of the map $f_T$.

# **Theories of Complexity**

## **Chaos theory**

The knowledge of science enables human beings to interpret various aspects and events that take place in nature. The interpretation is in mostly done by the use of models, especially the mathematical models. Therefore, equations which are constructed by the mind of a human are seen to be sufficient representation of the reality of nature. However, since the models follow mathematical logic, they are indisputable and non-contradictory. Scientist, therefore, has the responsibility of understanding whether the world events follow a laid down mathematical laws.

In constructing mathematical models, there are three major and successive steps that must be followed. The first step involves the observation of the existing phenomenon; in the second stage, the observed idea is translated into mathematical equations. Lastly, these mathematical equations are computed to give a solution to the phenomenon. Scientists, especially in the biological and medical fields, emphasized on the first step (the observation step) to understand the phenomenon's components thereby creating a body of knowledge which can then be models and translated into equations which facilitates the understanding of the observed phenomenon.

The theory of Chaos is still a developing mathematical theory which empowers a series of phenomena to be described in the dynamic field, i.e., the effects of force on the motion of a various object. All dynamic concepts trace their origin from Isaac Newton, who develops the laws of motions among other physical laws that are existing in nature. In the application of mathematical theorems, one must note the validity of their hypothesis within the framework of the question at hand. The most common hypothesis in dynamics is that time and space are ever continuous (that is, there is an infinite number of points between two fixed points). However, this hypothesis is rendered invalid in the cognitive neuroscience of perception since a finite time threshold is often considered.

The discovery of Chaos theory is credited to Edward Lorenz of MIT, in his adept in the weather prediction; he made calculation rounding with 3-digit instead of 6-digit numbers which did not prove the same solution. Moreover, in nonlinear systems which are explored through iterative mathematical processes, multiplications results to increase differences exponentially. Later Mitchell Jay Feigenbaum proposed what was later termed period doubling (on the logistic map concept) in describing the change from regular dynamics to chaos and vice versa. The logistic map is rather a function of [0,1] segment defined by:

$$X_{n+1} = r\, x_n\, (1-x_n)$$

Where: $n = 0, 1, \ldots$ The value n designates the time in discrete form of a single dynamical variable.

$$0 \leq r \leq 4 \text{ is a parameter}$$

It is observed that the dynamics of such a function shows very varied and different behaviors which are dependent on the values of r parameter. For instance,

Whenever $0 \leq r \leq 3$, the attractor of the system are of fixed point nature. The instability of attractor happens when r = 3. Whenever $3 < r < 3.57\ldots$ a periodic orbit is exhibited by the function as the attractor.where n is an integer. As r tends towards the value 3.57..., the function tends towards infinity.

When r is exactly 3.57..., the function has a Feigenbaum fractal attractor. And when r is exactly equal to 4, the function extends beyond the interval [0,1].

The above information can be represented diagrammatically as below:

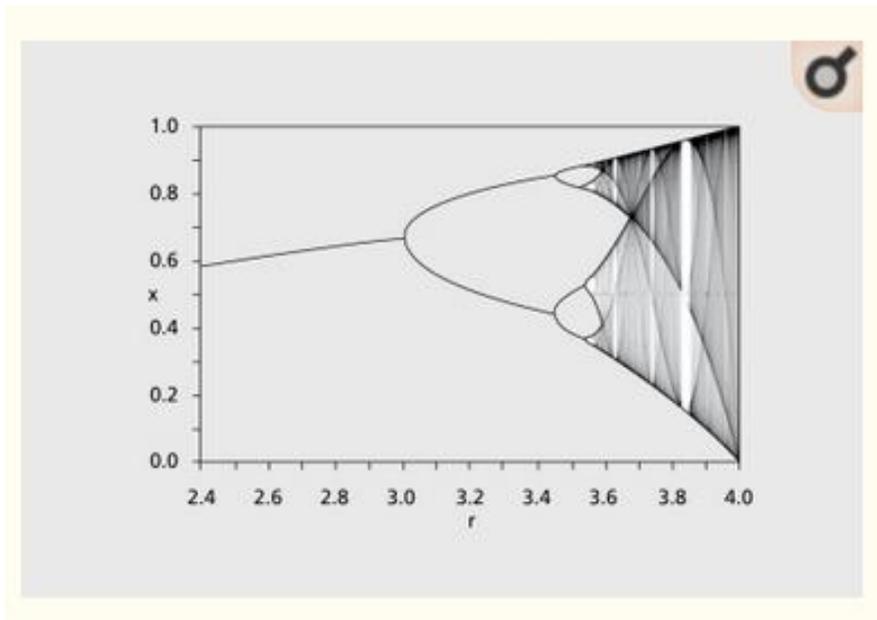

**Fig.1.** Showing how the parameter r (horizontal axis) varies with long-term x (vertical axis) [6].

The mathematical concept finds its application in various fields in human life such as determining the population of a locality where we consider the initial number of the individuals only. The r parameter can be taken as growth parameter or simply the birth rate.

Poincaré did research and showed that some dynamically nonlinear systems have behaviors which are either difficult to predict or totally unpredictable. This contributed to affirming the chaos theory. In the current world, chaos theory is not limited to mathematicians and scientists, who have advanced knowledge in mathematics, but it found its way into economics, social science, and the media. The success of chaotic theory have a wide bearing, artificial intelligence, deterministic terms, equilibrium, strange attractors, and unpredictability which are at the core of chaos theory have found their application in other fields where they are accorded various meanings depending on the context [7]. Thus chaos theory finds acceptance in the present social and philosophical preoccupations. The present day physics and mathematics and other scientific fields are so working extra miles to be so specific, and accurate in their undertakings and theories and thus makes use of chaos, entropy, and fractal dimensions, which gives a sense of reality to phenomena in the way they are perceived and measured.

**Chaos theory and medicine**

Discoveries in the medical field seem to infer that body organs function in a linear and deterministic mode. Example of the processes include the variation in the blood pressure and the diameter of the arteries, veins, and capillaries. However, the fact is that body organs function through coordination, which are under inherently complex laws. In affirming assumptions in the medical field, there is a need for having a large number of data points. The possibility of describing a biotic phenomenon in its right state space depends on whether the data points can be achieved or not. The use of nonlinear dynamical systems methods are useful in such study. A nonlinear dynamic system, however, can contribute to the understanding of various organ functions such as the respiratory system, cardiovascular system, and the brain.

In a research conducted by [6], reveals that the cardiac rhythm is very sensitive to both initial conditions as well as to its attractor's fractal dimension. Similarly, when the heart rate is so steady, it becomes very difficult for the heart to cope with changes in oxygen and blood circulation demands. Therefore it is clear that nonlinearity is required for the proper functioning of the vital body organs such as the heart and the brain. All these findings gives an insight on the role chaos theory play in the field of medicine and biology in general. In general, mathematical models play a very important role as they are used in describing rhythms. Thus the success of the chaos theory is attributed to epistemology.

**Thermodynamic principles applied to Complexity**

None-equilibrium thermodynamics when constrained by the second law of thermodynamics, results in the emergence of structures and processes in a class of dissipative systems. This law implies that if a system is moved away from the equilibrium condition, processes can emerge within the system so as to reduce the effect caused by the applied gradient [8]. If the dynamic condition is permitted within the system, the system is expected to organize itself. As living organisms undergoes growth and development in the ecosystem, it is expected by nature that they should increase their total dissipation, thereby developing more complex structures with have more energy flow rates. As a result, these living organisms will develop greater diversity, increase their cycling activity as well as generating more hierarchical levels.

Schrodinger suggested that life is distant from balance system and thus studying bodily systems from a non-equilibrium perspective would be necessary in reconciling thermodynamics and biological self-organization. Thermodynamics is the application of all energy systems such as chemical kinetic systems, classic temperature volume, and pressure system and, electromagnetic and quantum systems [9]. Thermodynamics generally addresses the system behavior in three major situations: classical thermodynamics which looks at a system at or near thermodynamic equilibrium, systems which are at some distance from the equilibrium and thus will try returning to the equilibrium position, and lastly, systems which have been detached from the equilibrium state and are constrained by some thermodynamic gradients so that the system may maintain given distance from the equilibrium position.

The first law of thermodynamics focuses on the relationship between work and heat and generally stated that energy can neither be created nor destroyed but can be transformed from one form to another and that within a closed system, the total energy remains the same. However, it has been found that the energy content of an enclosed system may change. Here the algebraic sum of heat and work is zero, i.e., heat and work are equal, and a state of thermodynamic equilibrium is achieved [10]. The second law of thermodynamics considers a process that is in action and requires that if there is any proceeding process, the energy quality in such a system must degrade. Irreversibility, i.e. entropy is also an acceptabl and excellent method of describing the first law of thermodynamics.

Scientific research has shown that dissipative structures whih are self-organize through the processes of fluctuations and instabilities may result in irreversible bifurcations and new stable system states [9]. However, such dissipative structures are stable only over a finite range of conditions. Dissipative structures are also found to be sensitive to fluxes and flow from outside the system [8]. The thermodynamic relationships hereto can be represented though coupled with nonlinear relationships. The non-equilibrium dissipative structure which exhibits coherent behavior includes autocatalytic chemical reactions, hurricanes, convection cells, and living systems. As the high order complex systems arise in the ecosystem, their growth and development occurs at the expense of entropy level increase in the system's hierarchy. The behavior is seen universally in both chemical and physical systems. This contributes to nonlinearity in the relationships among the systems in the ecosystem. A proper application of thermodynamic principles will help in the understanding of how each of the systems will behave in order to bring the ecosystem to equilibrium.

In reality, species which survive in the ecosystem are those are able to fully utilize the available energy in their development. The more process reaction of material and energy within a system, such as cycling, metabolism, and building higher trophic levels, the more entropy degradation or production is possible [10]. Thermodynamics helps in bringing and integrating these sources of energy in the ecosystem with the ability to utilize such energy. Nevertheless, the passage of the ecosystem from one steady state to another can be inversely related to the level of free energy production and utilization [8].

Systems which are nonlinear and non-equilibrium are usually open systems and therefore exchange both energy and matter with their surroundings. In so doing, these systems move from a lesser order to more order state. This process, however, decreases the system's entropy without going against the second law of thermodynamics [8]. This is because the sum of the total system entropy and that of its environment increases as required by the second law. This phenomenon is usually termed "out of order chaos," and the system which produces such phenomenon is termed "dissipative systems." In the current world, the study of chaotic, complex, and self-organizing systems (animal populations, meteorology, mechanics, and hydrodynamics, etc.) is in the rise [10]. Dissipative systems exhibits an incredible sensitivity to their surroundings, which can be better explained with the famous "butterfly effect."

Chaotic, complex, and self-organizing systems usually appear entirely random even though their operation is governed by deterministic equations [9]. Therefore, a grouping of phenomenologically random data which are constrained completely by deterministic equation forms the chaotic randomness. There are two main contributions of chaotic randomness in the ecosystem. First, it permits us to continously include into the deterministic framework of classical science broad areas of phenomena which are seen to be resisting such inclusion. Secondly, it is a representation of a limiting factor to the full stability of the classical paradigm [9]. This is owed to the fact that practical limitations on certainty brings to view our inability to accurately exclude the possibility that certain macroscopic phenomena may be openly random and therefore not a subject to deterministic equation's governance.

In conclusion, it is clear that biological complexity came into existence in conformity with thermodynamic principles. This fact is explained by the chaos, self-organization, and complexity theory. Thermodynamics helps in understanding how the energy in the ecosystem is utilized, but the nonlinear and chaotic systems.

# Mathematical Tools for Measuring Complexity

In these section, Poincaré plot, Shannon entropy, Sample entropy, Approximate entropy, Multiscale entropy, Detrended fluctuation analysis, Reccurence plot, Correlation dimension, and determinism have been considered in detail.

## Poincaré Plot

This is a time series which is geometrically represented in a Cartesian plane. Poincaré plot is very important in analyzing nonlinear processes in the body. For example, the Poincaré plot is used in the analysis and qualitative visualization of physiological signal and patterns such as that of heart rate dynamics (variability) which results from nonlinear body processes [11]. The plot is a 2D plot which is constructed by plotting consecutive and subsequent points in a representation of R-R time series. The representation is made on either phase space or Cartesian plane [12]. A normal Poincaré plot takes an elliptical shape and is defined by two standard descriptors i.e. *SD*1 and *SD*2 which are used in the quantification of the geometry of Poincaré plot [13]. The descriptors represent minor and major axes of the ellipse.

The descriptors are also described in linear statistics terms. That is, the descriptors in their standard form are guardian to the distribution's visual inspection [12]. In situations where interval of the sample is small compared to the short time correlation length, the various short-time correlations primarily observeable [13]. The Poincaré plot's current R-R interval, is able to influence a maximum of 8 subsequent RR intervals [11].

## Standard Poincaré Plot Analysis

In this analysis, R-R interval time series signal has been used in the plotting of Poincaré Plot, which has been denoted by $RR_n$. We make an assumption that there are a finite number of R-R interval for the plot.

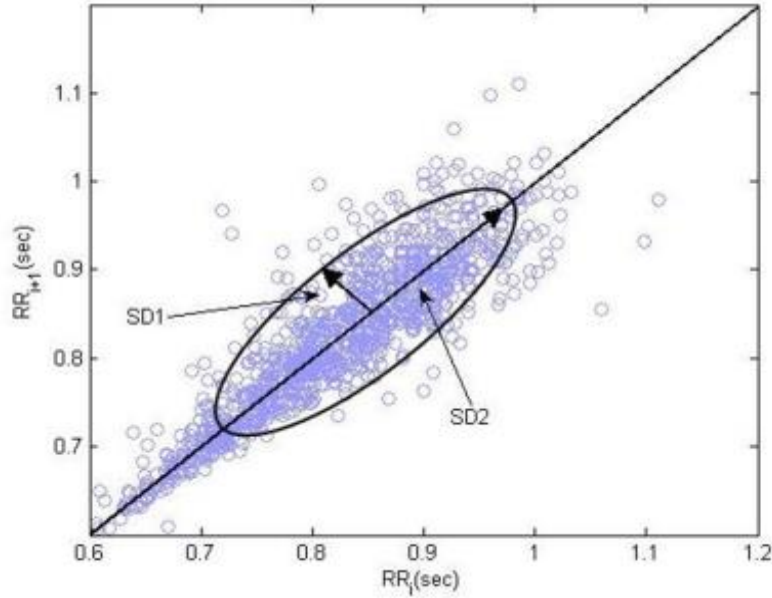

**Fig. 2.** RR intervals standard Poincaré plot of a healthy person. Adapted from [11].

From the above plot, $SD1$ and $SD2$ are the two basic descriptors, and the derivation of their mathematical origin can be achieved. In the diagram, it can be inferred the following:

1. The identity line is located at 45° imaginary diagonal line and that all the points on the line has a mathematical property of $RR_n = RR_{n+1}$

2. The measurement of dispersion of points which meet the line of identity at $90^0$ is achieved by the use of $SD1$ is used in, while the dispersion measured by $SD2$ is that which is along the line of identity.

Ideally, in a Poincaré plot $SD1$ and $SD2$ of have a direct relationship with basic statistical measures, the SD of RR interval (SDRR), and SD of the successive difference of R-R interval (SDSD) [14].

The relationship between these variables are given as:

$$SD1^2 = \frac{1}{2}SDSD^2 \qquad (1)$$
$$= \gamma_{RR}(0) - \gamma_{RR}(1)$$

$$SD2^2 = 2SDRR^2 - \frac{1}{2}SDSD^2 \qquad (2)$$
$$= \gamma_{RR}(0) + \gamma_{RR}(1) - 2\overline{RR}^2$$

Where $\gamma RR(0)$ and $\gamma RR(1)$ are termed the function of autocorrelation for lags 0 and 1 R-R interval and $\overline{RR}$ represent the intervals average. Considering the 1st and 2nd equations, it can be deduced that the derivatives of the correlation is the measure $SD1$ and the measure of $SD2$ and that the two measures forms the average of the R-R intervals time series. The two equations above are derivation of Poincaré plot with a unit time delay. For a better understanding of time series, there is a need for different time delays' plots [15]. When state space in two-dimensional view with lag-m is plotted, ther representation of $SD1$ and that of $SD2$ equations can be achieved as below equations:

$$SD1^2 = \gamma_{RR}(0) - \gamma_{RR}(m) \qquad (3)$$
$$\Rightarrow SD1 = F(\gamma_{RR}(0), \gamma_{RR}(m))$$

$$SD2^2 = \gamma_{RR}(0) + \gamma_{RR}(m) - 2\overline{RR}^2 \qquad (4)$$
$$\Rightarrow SD2 = F(\gamma_{RR}(0), \gamma_{RR}(m))$$

In the above cases of the 3rd and 4th equations, $\gamma RR(m)$ is viewed as the function of autocorrelation interval for $lag - m$.

## Complex Correlation Measure

The CCM measure evaluates the variation from point to point of the signal [11]. The CCM measure is usually a multiple lag function of correlation of the signal. Windowing for the perfect way of evaluating the CCM, which includes embedding the signal's temporal information [16]. Usually, during windowing, there are three Poincaré plot's consecutive points which are used. These three pints when joined to each other forms a triangle, the area of this triangle is computed. The cross-sectional area formed by the triangle has been established to be a measure of the variation temporarily of the windowed. When the pints are aligned in a straight line, it means that the area of the expected triangle is zero, this represents the co-linearity of the three points.

Furthermore, since three distinct points of the 2D plot are involved in each measure, the plot is therefore made up of not less than 4 distinct points of the time series for $lagm = 1$ and not more than six points whenever $lagm \geq 3$. Therefore, the measure gives four lag correlation information of the signal [14]. Taking a general case where we suppose that the i$^{th}$ window is made up of the three points; $a(x1, y1), b(x2, y2), c(x3, y3)$, the $i^{th}$ window triangular area (A) is calculated and can be obtained using the determinant:

$$A(i) = \frac{1}{2} \begin{vmatrix} x1 & y1 & 1 \\ x2 & y2 & 1 \\ x3 & y3 & 1 \end{vmatrix} \quad (5)$$

A is defined on R real line, and;

$A(i) = 0$, for a, b, and c points being on a straight line;

$A(i) > 0$, if there is counter clockwise orientations of a, b, and c points.

$A(i) < 0$, if there are clockwise orientations of the points a, b, and c.

For any *N* points forming a Poincaré plot, the CCM measure must be made up of all overlapping three-point windows and the CCM can be obtained from the relation;

$$CCM(m) = \frac{1}{C_n(N-2)} \sum_{i=1}^{N-2} \| A(i) \|$$

Where;

*m* is the lag of the Poincaré plot.

$C_n$ represent the constant of normalization.

$C_n = \pi \times SD1 \times SD2$; [17].

**<u>Shannon Entropy</u>**

Entropy is a familiar concept in thermodynamics, information theory, and statistical mechanics. This section looks into the formulation of entropy based on the information theory. The information entropy is usually termed Shannon's entropy about Claude E. Shannon, who was instrumental in the formulation of many of the key aspects of Shannon's information theory [18]. The theory came up at a time when the notion of information got precise and quantifiable. From a physical viewpoint, Shannon's information theory has no relationship with physics. However, Shannon entropy concept shares many intuitions with Boltzmann's, and various concepts of mathematics developed in information theory turn out to have relevance in thermodynamics and statistical mechanics [19].

Shannon information entropy, in its intuitive understanding, relates to the quantity of uncertainty of an event and the theory is thus concomitant with a given distribution. Shannon's entropy explanation shades more light on the uncertainty and defines entropy as a measurement of the mean content of the information linked with a random result [20]. This definition makes the intuitive distinction mathematically precise. The definition satisfies the following desiderate:

1. The entropy measure is supposed to be continuous, that is, a small change in any of the probabilities value should result in a small change in the entropy value.

2. Entropy is maximal if all the outcomes are equally likely. However, in many cases, the entropy change varies directly with the number of outcomes [18].

3. If there is certainty in the outcome, then the entropy should be taken as zero.

## The mathematical aspect of Shannon Entropy

Entropy according to Shannon is defines based on a discrete random variable *X*, with probable outcomes or states $x_1, ..., x_n$ as;

$$H(x) = \sum_{i=1}^{n} p(x_i) \log_2\left(\frac{1}{p(x_i)}\right) = \sum_{i=1}^{n} p(x_i) \log_2 p(x_i)$$

Where; $p(x_i) = \Pr(X = x)$ *is the probability of the $i^{th}$ outcome of the variable X*.

The above relation *means that the entropy of the variable X is the* arithmetic sum, divided by outcomes *xi* of X which are possible, of the creation of the likelihood of outcome *xi* multiplied by the logarithm of the reciprocal of the likelihood of *xi* (this is usually termed *xi's* surprisal). This concept is not limited to Shannon entropy but to a general distribution, as opposed to a discrete-valued event [20].

Shannon was able to prove that all entropy must be of the form:

$$-K \sum_{i=1}^{n} p(x_i) \log_2 p(x_i)$$

Where K is a constant. This is the form which satisfied his argument.

Shannon's definition of entropy can be used in an information source. When this is the case, Shannon entropy is applicable in determining the minimum network capacity required for reliable transmission of the source as encoded binary digits. The formula for calculating the minimum network capacity can be obtained by computing the possible outcomes of the data amount in any digital signal from the source of information.

**Relationship between Shannon Entropy and Thermodynamic Entropy**

There is a close and indisputable relationship that exists between thermodynamic entropy and Shannon entropy. The definition of Shannon's entropy, as outline in physical science and chemical engineering, shares a close relationship to that of thermodynamic entropy. Gibbs and Boltzmann are notable individuals who did substantial research in the field of statistical thermodynamics. Their works became a source of inspiration in adopting the term entropy in information theory which is now termed Shannon entropy. According to Jaynes (1957) which is taken from [20], thermodynamics as we know it should be considered as a real-time and practical use of information theory according to Shannon: The irreversibility found in thermodynamic is thus being interpreted as an estimation of the quantity of Shannon information (necessary in defining to details the minute state of a thermodynamic system) that is still unanswered by a explanation in the concept of the macroscopic variables of classical thermodynamics.

When we consider a heat addition process in a system, there is an obvious increase in the thermodynamic irrersibility since there is an increase in the quantity of possible microscopic states that can be occupied. This lengthens any description or explanation that can be made on a complete state [18]. A close consideration of Maxwell's demon reveals a hypothetical reduction in the thermodynamic entropy of a system based on the information on the states and the characteristics of the individual molecules.

**Approximate entropy**

This is the most common methods used in the quantification of the complexity of a given system [21]. Approximate entropy has been used as a means of quantifying the irregularity in time series data motivated by its applications to rather short, noisy data sets. In the recent past, entropy methods have been the primary methods employed when it comes to defining regularity or periodicity in human data has become quite common [22]. Mathematically, approximate entropy (abbreviated as ApEn) is part of universal development as the rate of entropy for an estimating Markov chain to a process.

The existence of repetitive array of pattern of variation in a time series renders approximate entropy more predictable than a time series in which such arrays are lacking. Approximate entropy reflects the probability that similar patterns of observations will not be succeeded by similar observations [23]. Therefore, any time series with many repetitive patterns have comparably small approximate entropy while more complex process which is actually less predictable has higher approximate entropy.

In the analysis of the approximate entropy, there are two important parameters, m, and r, where m represent the length of the runs while r is a filter. *Given N data points*, $\{u(i)\}$, form vector sequence ranging from x(1) to x (N-m+1). This can be defined by;

$$x(i) = [u(i), ..., u(i + m - 1)]$$

The above vectors are represention of m successive u values, starting with the i[th] point.

Defining the distance d, i.e. $d[x(i), x(j)]$, which is the distance between vectors $x(i)$ and $x(j)$ as the difference which can maximumly be achieved in their scalar components respectively. We use the order x(1), x(2), ..., x (N - m+1) in the construction, for each $i \leq N - m + 1$, $C_i^m(r) = $ (number of $j \leq $ N-m+1), such that:

$$d[x(i), x(j)] \leq \frac{r}{N - m + 1}$$

The $C_i^m(r)$'s measure within a tolerance r the frequency, or regularity of patterns which resembles window with length m pattern. We define the equation,

$$\Phi^m(r) = (N - m + 1)^{-1} \sum_{i=1}^{N-m+1} \ln C_i^m(r)$$

Such that ln represents the natural logarithm. Again we define the parameter,

$$ApEn(m, r) = \lim_{N \to \infty} [\Phi^m(r) - \Phi^{m+1}(r)]$$

When we have N data points, we estimate the above parameter by defining the statics;

$$ApEn(m, r, N) = \Phi^m(r) - \Phi^{m+1}(r)$$

From these equations, it is clear that entropy of approximation measures the probability that runs of patterns that are close for *m* observations remain close during comparisons. When there is a higher probability of continuing to be close, regularity gives smaller values of the entropy of approximation, and when there is less chances of rememaining and continual keeping close to one another, regularity produces large approximate entropy values [23]. Upon unraveling definitions, we deduce;

$$-ApEn = \Phi^{m+1}(r) - \Phi^{m}(r)$$

This is equal to the average over i of ln, which is the conditional probability that:

$$| u(j + m) – u(i + m) | \leq r, \text{ given that;}$$

$$| u(j + k) – u(i + k) | \leq r; \text{ for } k = 0, 1, …, m\text{-}1.$$

A better and higher intuitive physiological consideration of this definition requires the use of a multistep descriptive algorithm, which is embedded with a figure [22].

Limitations of Approximate Enthalpy

There are two major limitations of the approximate entropy:

1. This entropy depends entirely on the record length whci shoud not be the case. Again approximate entropy is averagely lower than what is expected in cases of short records.

2. Approximate Enthalpy lacks relative consistency. For instance, if the approximate enthalpy of a given data set is large compared to another data set, that entropy should remain larger for all conditions tested. However, this is not the usual case.

**Sample Entropy**

The most crucial task in understanding the mechanism governing the dynamics of a system is the ability to quantify a complex signal. Such quantification can only be achieved upon application of reliable methods [24]. Approximate entropy analysis is the most commonly used methods in quantifying the complexity of a given system. However, there are numerous pitfalls in the approximate entropy and therefore calls for application of the more advanced method in the analysis of complex systems [25]. One such method is the use of sample entropy. Sample entropy (commonly abbreviated as SampEn) is seen as a adjustment of approximate entropy. It is primarily used for evaluating the complexity of an of physiological time-series signals, and diagnosing diseased states. When compared to approximate entropy, sample entropy has two main advantages; data set independence and ease of implementation [21].

In the analysis using Sample entropy, we consider a data set a set of data x(i), where i ranges from 1 to N, sampled at a discrete time interval Δt. It should be noted that for discrete systems such as stochastic process (e.g., white noise) and RR interval among others, Δt = 1. However, for continuous systems, Δt is taken as the inverse of the sampling frequency of the process [22]. In sample entropy analysis, x(i) is divided into vectors (overlapping blocks) with dimension (size) m. The overlapping gap has a natural time gap of one sample unit in between any two successive components of the block. The procedure used in sample entropy analysis is similar to that applied in the reconstruction of an attractor from time series [23].

Now, the $i^{th}$ m-dimensional block V(i) is represented as;

$$V(i) = \{x(i), x(i + \delta), ..., x(i + [m - 1] \delta)\},$$

Where;

$\delta$ = successive components' time delay of the vector. In complexity analysis, $\delta$ = 1.

The reconstructed vectors are the state space (of dimension m) representation of the dynamics of a system. Sample entropy can thus be viewed as the logarithmic difference between the probability density that the vector V(i) will occur within a chosen distance r in m-dimension and the probability that the vector V(i) will occur within the same chosen distance r in (m + 1)-dimension. Determining the density is usually a challenge. However, the density can be estimated, in estimating the density (represented hereafter as $\rho^m(r)$), of the state space with m-dimensional [21].

The procedure for calculating the radius is repeated for all vectors. The procedure is again repeated for the (m + 1) state space vectors reconstruction for the assessment of the density of $\rho^{m+1}(r)$ in $(m+1)$ dimension. The measure of the fraction of the reconstructed vectors forms the definition of state space density. However, the vector must fall within the chosen radius [21]. Therefore, mathematically Sample Entropy can be defined as;

$$SaEn(m, r) = \log[\rho^m(r)/\rho^{m+1}(r)].$$

While carrying out this process, it must be ensured that the densities as $\rho^m(r)$ and $\rho^{m+1}(r)$ are estimated for the same number of vectors to avoid the bias in the estimation.

In some cases, we apply the Grassberger–Procaccia integrals for the estimation of these densities [24]. Many times, it is difficult to establish the relationship between dynamical correlations and dynamics (such relationship may not be existing). For instance, when two different frequencies are used to sample the same signal. At times, dynamical correlations are associated with the system's intrinsic dynamics. This association, in many cases, do not allow the accurate quantification of how complex a given system is. In overcoming the difficulties associated with such correlations, δ (which represent a delay time) is introduced between the the block's subsequent components. The delay time δ is chosen to represent that point of time where thefunction of autocorrelation falls below 1/e [22].

Time delay δ is greater than unity in the case of dynamical correlations. When the time delay is considered, Sample Entropy can be expressed as;

$$SaEn(m, r, \delta) = (1/\delta) \log[\rho^m(r)/\rho^{m+1}(r)].$$

In the density approximation or each center i, we discard (i + δ) number of points. This help in accounting for the temporary correlated points. Sample entropy finds application in the analysis of heartbeat (RR) interval, congestive heart failure (CHF), deterministic systems, and stochastic processes.

**Multiscale Entropy**

As was mentioned in the approximate entropy section, ApEn is the most common method employed to analyze the system'scomplexity. However, approximate entropy is associated with some methodological pitfalls, and thus, there is a need for the development of other methods. In order to overcome the methodological pitfalls in the ApEn, sample entropy was developed. However, in addition to correcting some of the biases in the ApEn, the current definition of sample entropy is not sufficient to completely characterize the complexity of a process or a system [26]. This led to the introduction of the Multiscale entropy (abbreviated as MSE). When the system is approached by the use of the multiscale entropy, it is important that the signal be subdivided into a given number of disjoint windows with each signal having the size $\tau$. It is equally important to note that averaging of the data must be done inside each time window. Multiscale entropy, when vied as a variation of sample entropy, can be considered as a function involving $\tau$ as one of the variables. From the perspective of signal processing, obtaining the data average inside a time window $\tau$ is considered to be equal to performing down-sampling by a factor of $\tau$ on the data which leads to the reduction of the high-frequency components of the signal. Similarly, the averaging procedure used in such cases is equivalent to using the low pass filter with $\tau$ determining the filter cut-off [26].

Biological systems form an example of highly integrated and complex systems which functions at multiple time scales. For instance, neurons functioning, Circadian rhythms, reproductive cycles, and bone remodeling, all these operate at different time scale [27]. This implies that bio signals are multiscale signals and depending on the scale on which these signals are examined, they are likely to exhibit different behaviors such as nonlinearity, extreme variations, non-stationarity, long memory, and sensitive dependence on small disturbances among others [28]. In the analysis of such multi-scaled signals, the use of Multiscale entropy is indisputable. MSE has found application in many fields which involve the operations of systems with multiple time series such as in the heart rate monitoring, EEG dynamical complexity assessment, and the analysis of blood flow which characterize psychological dimensions in non-pathological subjects [27].

Systems which are diseased, if related with the development of more regular behavior, displays results of lower entropy values when matched with the results obtained from dynamics of free-running healthy systems. Some pathologies are concomitant with high variations with statistical features resembling the properties of uncorrelated noise [29]. Algorithms of traditional format produces entropy values' increament for noisy, pathologic signals when matched with healthy dynamics which are showing related (1/*f*-type) properties, even though healthy dynamics characterize a more physiologically complex, adaptive states.

Mathematically, given a 1D time series of discrete nature of the form

$$\{ x_1, ..., x_i, ...x_N \},$$

A coarse-grained series of time of the form $\{y^{(\tau)}\}$ must be consecutively constructed. The time series is dependent on τ (i.e. the series scale factor) as depicted in the below equation;

$$y_j^{(\tau)} = \frac{1}{\tau} \sum_{i=(j-1)\tau+1}^{j\tau} x_i, \quad 1 \leq j \leq N/\tau$$

For scale 1, the time series remain constant i.e. it is the original time series equals the new series. Again, the time series of a coarse-grained nature whose length equals the quotient when the length of the initial time series is divided by τ (time series scale factor). For instance, when one has a time series with 30000 points, he course-grain them up to scale 20 such that the shortest series contains 1500 points. With this information, we can computre the measure of sample entropy for each of the time series of coarse-grained nature plotted as a scale factor τ function. The procedure of obtaining SampEn of each time series and plotting the series as a function of scale factor forms the Multiscale entropy (MSE) measure [26].

The next step is to carry out a test on the multiscale entropy method on virtual white and i/f noises. A research conducted by [26] for scale 1, showed that a higher value of entropy should be assigned the white noise time series as compared to 1/*f* time series. The value of enthalpy measured using all the scales is found to remains constant while the enthalpy value for the time series of white noise nature is observed to be decreasing [29]. The white noise time series decrease such that for scale less than 5, it is smaller than 1/f noise. This is because complex structures which occur across multiple time scales constitute 1/*f* noise.

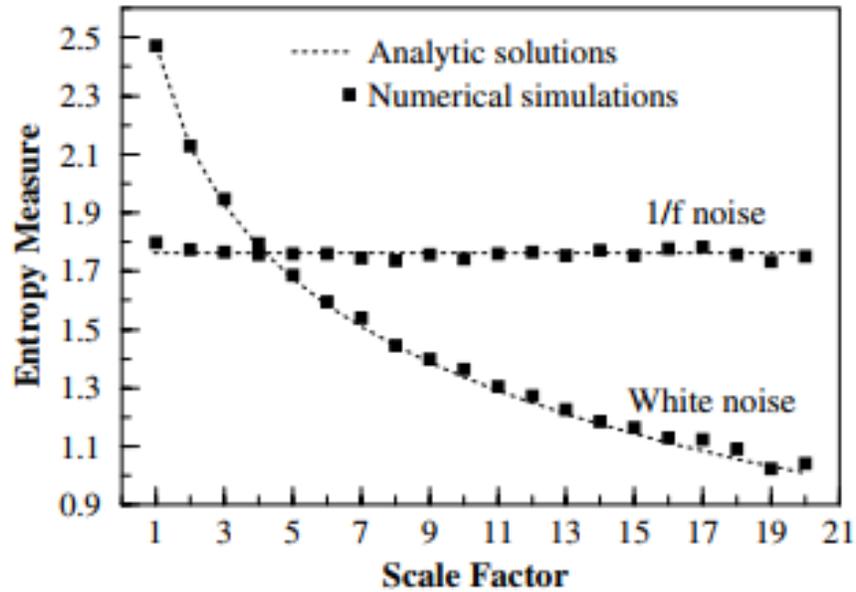

**Fig.3.** Gaussian MSE analysis distributed white noise and 1/f noise [26].

The next step involves the application of Multiscale entropy method in the analysis of sampled physiologic time series. The Multiscale algorithm gives steady and reliable findings when applied in the assessment of complexity of both simulated correlated and uncorrelated noises and the integrated result of a physiologic control system [28].

**Detrended Fluctuation Analysis**

In chaos theory, time series analysis, and stochastic, Detrended Fluctuation Analysis (commonly abbreviated as DFA) is considered a method for determining the signals' statistical self-affinity. DFA is primarily useful when carrying out time series analysis, which appear to be long-memory processes or 1/f noise [30]. DFA may as well be applied in signals analysis (i.e., mean, mode, variance, standard deviation, etc.) and dynamics are non-stationary, i.e. changing with time. This approach is related, in its application, to autocorrelation and Fourier transforms [31]. DFA has successfully used for analysis in diverse fields such as DNA sequence, neuron spiking, long-time weather records, and heart rate variability, among others. The main reason behind the employment of DFA in such filed is that it helps in avoiding spurious detection of correlations which are the objects of non-stationarities in the time series [32].

Mathematically, the DFA analysis method is consisting of five major steps. Each of these methods must be strictly followed if a correct result is to be realized.

Supposing $x_k$ is a length N series, it is said that the series is of compact support, i.e., $x_k = 0$ for any fraction of the values which is insignificant only;

Step 1: definition of the profile

Let the equation of the profile be;

$$Y(i) \equiv \sum_{k=1}^{i} [x_k - \langle x \rangle], \qquad i = 1, \ldots, N. \qquad (1)$$

It is not mandatory that the mean of x be subtracted in the first step since the third step will eliminate it through the detrending method.

Step 2: Division of the profile.

In this step, the profile Y(i) is sub-divided into non-overlapping segments of $N_s$ = int (N/s) form. The segments' lengths s are the same. Usually, a short part of this profile do remain because the length N of the series is not always a multiple of the s time of scale which is under consideration. To avoid disregarding this portion of the series, a similar procedure is followed but now it starts from the opposite end thus 2Ns segments are obtained.

Step 3: Calculation of the local trend.

In this step, each of the 2Ns segments' local trend is calculate using a least-square fit of the series. Then we consider for the trend the variance. The determination of the variance of the trend is achieved using the subsequent equations.

$$F^2(s,\nu) \equiv \frac{1}{s}\sum_{i=1}^{s}\{Y[(\nu-1)s+i] - y_\nu(i)\}^2 \qquad (2)$$

For each segment v, v = 1, ..., Ns and

$$F^2(s,\nu) \equiv \frac{1}{s}\sum_{i=1}^{s}\{Y[N-(\nu-N_s)s+i] - y_\nu(i)\}^2 \qquad (3)$$

For v = Ns +1, ..., 2Ns.

In the above scenario, $y_v(i)$ is the segment v's fitting polynomial. Linear, cubic, quadratic, quartic, or any other polynomial of higher order can also be applied in the procedure of fitting the segment.

Step 4: Segments Averaging

We average all the segments thereby obtaining the $q^{th}$ order function of fluctuating nature as;

$$F_q(s) \equiv \left\{\frac{1}{2N_s}\sum_{\nu=1}^{2N_s}[F^2(s,\nu)]^{q/2}\right\}^{1/q}, \qquad (4)$$

However, q which is the index variable may assume all real value apart from zero [32].

Step 5: Scale Behavior Determination.

By analyzing, for each value of q, the log-log plots Fq(s) against s, we are actually determining the behavior of scaling functions of fluctuating nature. If the xi series are long-range power-law correlated, then Fq(s) is seen to be increasing, for large values of s, as a power-law. This can mathematically be represented as;

$$F_q(s) \sim s^{h(q)}. \qquad (5)$$

Generally, the exponent h(q) may depend on q. However, for mono-fractal time series with compact support, h(q) is assumed to be independent of q, this is because the scaling characteristic of the variances $F^2(s, v)$ is the same for all the v segments.

The above procedure gives a summary of the procedure of carrying out Detrended Fluctuation analysis. In summary, it should be understood that since DFA is dependent upon line fitting, it is always possible to find a scaling exponent by the DFA method. However, this does not necessarily imply that the time series is self -similar [30].

**Recurrence Plots**

Recurrence is defined as a vital property of dynamical systems, that can be utilized in characterizing the behavior of the system in phase space. Recurrence plot is a powerful tool for visualization and analysis [33]. For theories involving chaos, this plot (hereafter abbreviated as RP) is defined as a plot that shows the relationship between the times when the trajectory visits approximately the similar path in the state space as at the time *I* with each moment i in time. Mathematically, the information above can be represented as;

$$\vec{x}(i) \approx \vec{x}(j),$$

A recurrence is described as the average period of time the trajectory takes to returns to a location it had pass through, i.e., the period. Recurrence plot therefore shows (expresses) the pairs of times collection when the trajectory was at a similar location it had previously visited , i.e., the set of (is, js) with

$$\vec{x}(i) = \vec{x}(j).$$

From this definition, the following can be deduced; when any of the trajectory is exactly periodic with periodic time T, it implies that all the times pairs will be separated with a multiple of the periodic time T. these, therefore, will be observed as diagonal lines. The plot can only be made using discrete times and discrete phase space; therefore, in case there are continuous times and continuous phase space; there must be discretized. For instance, if we take a vector x(i) as the location of the trajectory at times $i_\tau$ [34]. When we consider any time the trajectory gets very close (i.e., with ε) as a recurrence, then recurrence or non-recurrence can be represented using the binary functions as;

$$R(i,j) = \begin{cases} 1 & \text{if } \|\vec{x}(i) - \vec{x}(j)\| \leq \varepsilon \\ 0 & \text{otherwise,} \end{cases}$$

And if $R(i,j) = 1$, the recurrence plot is likely to put a point at the (i, j) coordinates.

**Structures in Recurrence Plots**

The recurrence plots exhibit the characteristics of both the small scale and large scale patterns. The patterns are denoted as typologies or texture. The typologies can be characterized as homogenous, periodic, drift, and disrupted. Homogenous are of stationary and autonomous systems and has a short relaxation time. Periodic characteristic is possessed by those systems which are oscillating. Drift is as a result of systems with slowly varying parameters. And lastly, the abrupt changes in the dynamics are as a result of bands (white areas) in the recurrence plot.

These four typologies are as outlined below.

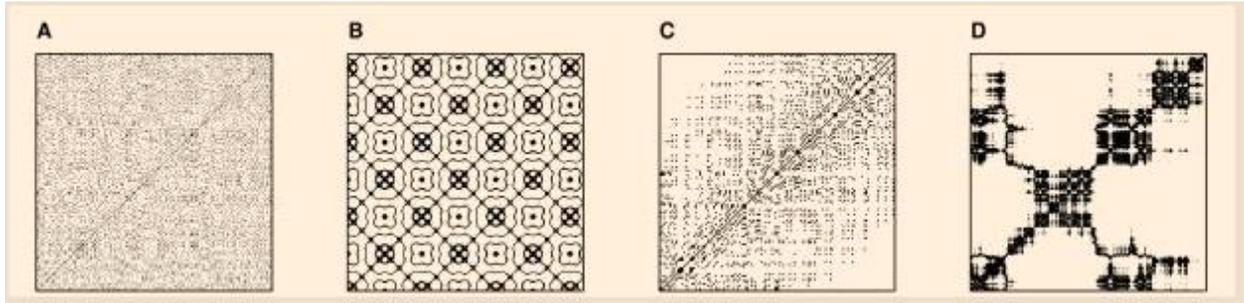

**Fig.4.** Typologies (A –homogenous, B-periodic, C-drift, and D-disrupted) in RP [35].

To contain many variables, recurrent plots can be extended. Such extensions of the plot are referred to as multivariate of the recurrent plot. There are two types of multivariate extensions, cross recurrence plot and joint recurrence plot [35].

The cross recurrence plots look into the trajectories in phase space of two different systems which are in the same state space. These can be represented mathematically using the equation;

$$\mathbf{CR}(i,j) = \Theta(\varepsilon - \|\vec{x}(i) - \vec{y}(j)\|), \quad \vec{x}(i), \vec{y}(i) \in \mathbb{R}^m, \quad i = 1, \ldots, N_x, \; j = 1, \ldots, N_y.$$

While using the cross recurrence plot, the two systems must have same dimension. However, the data length may differ. These plots are used when comparing the occurrence of two systems with similar states. The plots may also be applied in the analysis of the dynamical evolution similarities of two different systems.

There exist another form of reccurence plot which is known as the Joint recurrence plots. These plots are considered to be the Hadamard product of the recurrence plots of the sub-systems under consideration [34]. For instance, the joint recurrence plot for any two systems x and y, can be defined as;

$$\mathbf{JR}(i,j) = \Theta(\varepsilon_x - \|\vec{x}(i) - \vec{x}(j)\|) \cdot \Theta(\varepsilon_y - \|\vec{y}(i) - \vec{y}(j)\|), \quad \vec{x}(i) \in \mathbb{R}^m, \quad \vec{y}(i) \in \mathbb{R}^n, \quad i,j = 1, \ldots, N_{x,y}.$$

While crosses recurrence plot compares the occurrence of similar states of two systems, the simultaneous occurrence of *recurrences* in two (or more) systems comparision is established by the used of joint recurrence plots. Even though the dimension of the phase space under consideration can be the difference, states for the sub-systems being considered, however, must have equal number [35].

**<u>Correlation Dimension</u>**

In chaos theory and nonlinearity, the correlation dimension (usually denoted by v) is defined as a measure of the dimensionality of the space occupied by a set of random points. For instance, given a set of random points on the real number on the range [0, 1], the correlation dimension will be $v = 1$. However, the set of random point is distributed, for instance, a triangle embedded in 3D space (or any dimensional space such as m-dimensional space. The correlation dimension will be $v = 2$ [36]. The real field where the correlation dimension is utilized is in the determination of (mainly fractional) dimensions of fractal objects [37].

There other methods for measuring dimensions such as the utilization of the box-counting dimension, the Hausdorff dimension, and the information dimension. However, correlation coefficient dimension is preferred due to its straightforwardness and faster way through which it is calculated; correlation coefficient is by far less noisy in cases where only a small number of points is available for consideration to compare to the methods mentioned above [38]. The method is also agrees, in most cases, with other calculations of dimension. The dimension measurements are usually used in the quantification of the space filling properties of a set of points. It must be noted that space where the set is embedded, which also has an associated dimension, can be a real space or an abstract mathematical space [36].

Mathematically, a set of N points in m-dimensional space can be represented as;

$$\vec{x}(i) = [x_1(i), x_2(i), \ldots, x_m(i)], \qquad i = 1, 2, \ldots N$$

Asrepresented above, the correlation integral, denoted as C(ε), is obtained from the relation

$$C(\varepsilon) = \lim_{N \to \infty} \frac{g}{N^2}$$

Where N = the points' total number which are in the m-dimensional space,

g = total number of pairs of points having a distance less than ε between them.

From the relation it can clearly be depicted that as the number of points increases (i.e., tends to infinity), and the distance between these points tends decreases (i.e., tend to zero), the correlation, especially for a small value ε, eventually takes the form;

$$C(\varepsilon) \sim \varepsilon^{\nu}$$

Correlation dimension can also be obtained using a graphical method. In the application of the graphical method, there are two conditions that the data set must meet, the number of points must be sufficiently large, and the points must be evenly distributed [38]. If such circumstances are met, the log-log of the correlation integral against ε gives a proper estimate of the correlation dimension. The idea is basically based on the facts that when we have higher dimensional objects, there are numerous ways through which points can be close to each other and therefore, the number of pairs close to each other will rise more rapidly for higher dimensions [36].

**Determinism**

Determinism finds its application in many fields such as in the computer technology where it plays a role since it is employed in the determination of various computer systems' complexity of [39], the analysis of river flow dynamics [40], and in the analysis of various complex body systems [41]. Determinism detection and nonlinearity detection of systems in time series are some of the example of fields which have received more attention in the time series analysis of nonlinearity. Determinism method needs to be complemented with other methods of analysis since it is very difficult to obtain from a natural phenomena, reliable time series clear-cut results.

In testing for nonlinearity and determinism, we apply the univariate approach or method of analyzing time series which is nonlinear. Such methods employed in the analysis include the method of reconstructing state-space which is a modern approaches in the analysis of nonlinearity determinism[42]. Since in the approach which involves univariates, each variable is considered to have the information on system dynamics overally, therefore it becomes easy when reconstructing embedding space which is are usually equivalent to the initial state space of the system from a time series of a single scalar nature [42]. Considering a time series $\{q_i\}_{i-1, ..., N}$, the sequence of M = N – (m - 1)τ vectors of m-dimension which is of the below form topologically describes an object which is equal to the system's attractor.

$$Q_i = \{q_i, q_{i-\tau}, q_{i-2\tau}, ..., q_{i-(m-1)\tau}\}$$

The dimension of the space denoted as m is the embedding dimension while the τ parameter is the suitable time delay [40]. It is therefore possible to describe the dynamics of the system in discrete map format which is present in the embedding space. Such dynamics of the system takes the form;

$$Q_{i+1} = f(Q_i)$$

**<u>Nonlinear redundancies in determinism</u>**

Redundancies have been employed in testing for non-linearity in systems. Suppose we have delay variables m, redundancies are therefore take to be the mutual information *m*-dimensional extension of the form:

$$R(m, \tau, \delta) = mH(m = 1, \tau = 0, \delta) - H(m, \tau, \delta)$$

Where, $H(m, \tau, \delta)$ is the entropy according to Shannon in m-dimensional extension. Therefore, such entropy forms a joint probability distribution of the variables,

$p(q_i, ..., q_{i-(m-1)\tau})$

That is contained in the

$\{ q_i, ..., q_{i-(m-1)\tau} \}$

Whenever we need to increase the exactness of such an estimate, we use the second order redundancies:

$$R_2(m,\tau,r) = -m\log_2[C_2(m=1,\tau=0,r)] - \log_2[C_2(m,\tau,r)]$$

and

$$R'_2(m,\tau,r) = -\log_2[C_2(m=1,\tau=0,r)] + \log_2\left[\frac{C_2(m,\tau,r)}{C_2(m-1,\tau,r)}\right]$$

The integral of the second-order correlation relates to these second order redundancies through the relational equation;

$$C_2(m,\tau,\delta) = \frac{1}{M}\sum_i^M \frac{1}{m-1}\sum_{j\neq i}^M \vartheta(r - \|Q_i - Q_j\|)$$

Where; v = the step function according to Heaviside:

v = 0 whenever zero is equal or less than the argument, and the function is equal to 1 otherwise.

| | is the vector norm, and $r$ = radius of a region about the generic $Q_i$.

Second order Kolmogorov-Sinai (KS) entropy is related to second order marginal redundancy by the following equation.

$$K_2 = \lim_{m\to\infty}\frac{R'_2(\tau=\tau_1) - R'_2(\tau=\tau_2)}{\tau_1 - \tau_2}$$

This is simply the average rate of information being created by the system [40]. Kolmogorov-Sinai entropy establishes the quantity of the arithmetic mean of new information concerning the system's complexity and dynamics. Measurement of the time series new values brings out such pieces of information. According to [42], it is needful for us to use Kolmogorov-Sinai to characterize the average certainty of a system. The system's average certainty usually represents the positive Lyapunov exponents summation [42].